# One-magnon (electromagnon) light scattering in BiFeO$_3$ single crystals

**Manoj K. Singh[1],  Ram S. Katiyar[1] and J. F. Scott[2]**

[1]*Department of Physics and Institute of Functional Nano Materials, University of Puerto Rico, PR 00931-3343, USA*
[2]*Department of Earth Science, University of Cambridge, Cambridge CB2 3EQ, U. K.*

We observed Raman scattering from magnon in frequency range from 10 to 65 cm$^{-1}$ in BiFeO$_3$ single crystals   at cryogenic temperatures; the temperature dependence of the magnon frequency at 18.2 cm$^{-1}$ approximates an $S = 5/2$  Brillouin function up to the temperature (~280 K) at which the magnon becomes overdamped. The diverging cross-section and the frequency-shift at ~140K and ~200 K implies a magnon-reorientation transition as in orthoferrites.  Magnons in polar materials such as BiFeO$_3$ are often termed "electromagnons," meaning that they possess an electric dipole moment due to magnetoelectric coupling.

Ferroelectromagnetic materials, i.e., multiferroics, exhibiting ferroelectric (or antiferro-electric) properties in combination with ferromagnetic (or antiferromagnetic) properties[1,2]. Additionally they exhibit the phenomenon called the magnetoelectric coupling; i.e. magnetization induced by electric field and electric polarization by a magnetic field. These phenomena also give rise to unusual dynamical effects, which can be observed in optical experiments. Magnetic excitations in multiferroic polar crystals [1,2]are generally not pure spin waves but contain significant contributions to their Raman scattering cross-sections from electric dipole matrix elements. These so-called electromagnons exhibit different dynamical characteristics from magnon scattering in centrosymmetric lattices. [3-6]

Bismuth ferrite has become an extremely popular material at present because its rare or even unique properties of having room temperature ferroelectric and magnetically order [7]. It is a rhombohedrally distorted ferroelectric perovskite ($T_c \approx$ 1100K) with the space group R3c. $BiFeO_3$ shows G-type canted antiferromagnetism up to 643K ($T_N$) [7], in which all neighboring magnetic spins are oriented antiparallel to each other. In addition, the axis along which the spins are aligned precesses throughout the crystal, resulting in a modulated spiral spin structure with a long periodicity of ~620Å . It has a rather small magnetization but very large polarization of ca. 100 $\mu C / cm^2$ , both in thin-film[8] and single-crystal form[9]. Despite this abundant research on $BiFeO_3$, relatively little is known about its spin waves. In the present study we grew highly pure single crystals by flux method and obtained Raman spectra of high resolution and excellent signal-to-noise down to 10 cm$^{-1}$. These reveal two strong one-magnon lines and four relatively week branches in between the frequency range 10 to 65 cm$^{-1}$, and we show that the lower frequency one follows the temperature dependence of an $Fe^{+3}$ S=5/2 Brillouin function up to 280 C (0.44 $T_N$), above which it becomes overdamped.

A single crystal of $BiFeO_3$ was prepared by employing a flux method. [10].The crystals display a rhombohedral (hexagonal) structure and the XRD pattern of single crystal is exclusively characterized by the [010]$_{cubic}$ or [012]$_{hex}$ [10]. Raman-scattering data were obtained using a T64000 spectrometer (Horiba Inc.) equipped with a triple-grating monochromator and a Coherent Innova 90C $Ar^+$-laser with the excitation wavelength at 514.5 nm. The spectral resolution was typically less than 1 $cm^{-1}$.

Raman-scattering data for BFO single crystal were obtained by employing two distinct normal backscattering VV and VH geometries ( systematically represented in inset of Fig. 1),  and their results at 80K are presented in Fig. 1.  In these configurations, the propagation direction of the relevant phonon wave vector **k** is parallel to the $[010]_c$ axis of cubic  BFO.  The spectra were from a face of a single crystal (010) that is not perpendicular to a principal axis of polarization, i.e. along (111), and therefore it includes modes of all symmetries. As presented in Fig. 1, we observed two intense peaks at 18.2 and 26.6 $cm^{-1}$ and weak-intensity frequency modes at 34.5, 43.4, 51.4 and 60.7 $cm^{-1}$ in both scattering configurations.   In addition, we also observed 14 first-order Raman peaks at 80K; these modes are comparable to the recently reported Raman spectra of $BiFeO_3$ with the same R3c symmetry [11,12]. The two intense one-magnon branches are at 18.2 and 26.6 $cm^{-1}$.  These rather large intensities are compatible with them being "electromagnons" with an electric dipole moment and not a pure spin wave. The lower-energy line agrees [13] with the sharp infrared reflectivity feature at 20 cm-1. Katsura et al [5] developed a theory for explaining the neutron scattering spectra in helical magnets and suggested that the two lowest modes are due to conventional magnon scattering origin whereas other modes arise magnon dispersion and a folded magnetic Brillouin zone and are hence relatively weak in magnitude. This proposed model is very close to $BiFeO_3$ experimental data, in which the observed one-magnon lines at 18.2 and 24.6 $cm^{-1}$.

The intensity of the first two magnon modes at 18.2 and 24.6 $cm^{-1}$change abruptly as we change the configuration from VV to VH in Fig 1 revealing that these two modes follow the polarization configuration similar to that observed from magnons in orthoferrites [14-16]. The variation of frequency shift and intensity of these two modes with temperature are also quite different from others modes above 30 $cm^{-1}$.

In order to study the origin of these two one magnon in $BiFeO_3$ single crystal we compared our result to the rare earth orthoferrites [14-16]. The classic study of Raman scattering from magnons in orthoferrites is that of White, Nemanich, and Herring [14].  They found in each of five isomorphs two one-magnon branches labeled $\sigma$ and $\gamma$. The $\sigma$ branch has polarizability tensor components xz and yz; whereas the $\gamma$ mode has xx, yy, zz, and xy. The $\gamma$ magnon mode is necessarily the higher in frequency. In ErFeO$_3$ [15], they found the room-temperature frequencies of these two modes to be 14 and 23 cm$^{-1}$; these values are very close to our results in BiFeO$_3$. These frequencies are given by

$$\omega_\sigma = 24JS[2(K_x - K_z)S + 2JSa^2k^2]^{1/2} \qquad (1)$$

$$\omega_\gamma = 24JS[6DS\tan\beta + 2K_xS + 2JSa^2k^2]^{1/2} \qquad (2)$$

Here J is the exchange energy between Fe-ions separated by a; D, the antisymmetric exchange; $\beta$, the weak canting angle; K$_{i,j}$ are anisotropy constants; and k, wave vector. The orthoferrites such as ErFeO$_3$ have orthorhombic crystallographic (chemical) primitive cells and also orthorhombic magnetic cells [15]. The crystallographic chemical cell in BiFeO$_3$ is rhombohedral; however, the magnetic cell is monoclinic with space group Bb[17], with a small distortion from orthorhombic. Therefore we can use Eqs.1, 2 for bismuth ferrite in its ambient phase.

Venugopalan et al. have pointed out [16] that the $\sigma$ magnon branch corresponds to cooperative motion of the magnetization vectors of the two sublattices along narrow elliptical orbits in the ac-plane of RFeO$_3$ ( R = Tb, Tm) single crystals; this magnon has the net ferromagnetic moment oscillate at the characteristic ferromagnetic resonance frequency. In contrast the $\gamma$ mode has magnetization vectors precess along elliptical orbits whose major axis is the b-direction, and this produces no change in ferromagnetic moment. This is then often termed the antiferromagnetic resonance mode. Whereas the $\sigma$-mode is very temperature dependent near a spin-reorientation transition temperature, the $\gamma$-mode is unaffected.

To study the behavior of observed electro-magnons with temperature in BFO single crystal, we performed micro-Raman scattering measurements by employing two different normal backscattering geometries (VV and VH) in the temperature range between 80 and 300 K, and their results are presented in Fig 2(a,b). In general, we observed that the $\sigma$-magnon at 18.2 cm$^{-1}$ associated with ferromagnetic ordering is strongly temperature dependent, whereas $\gamma$-magnon at 26.6 cm$^{-1}$ and other magnon frequencies at 34.5, 43.4, 51.4 and 60.7 cm$^{-1}$ are disappeared above certain temperature without showing any remarkable frequency shift.

Fig.3(a) illustrates the $\sigma$ magnon (18.2 cm$^{-1}$) scattering intensity from 80-280K. The singular peak at 140 and 200K is reproducible, and in agreement with the frequency data in Fig.3 may indicate some spin reorientation heretofore unpredicted. Fig. 3 (b) shows that the magnon line width remains very narrow (3.0-3.5 cm$^{-1}$) and nearly temperature independent from 80-230K.Fig. 3(c) shows that the $\sigma$ -magnon frequencies vary approximately as the Brillouin function for S=5/2 up to 280 K[22]. Note that the data points near to 140 and 200K do not fall on the curve. Above 280K the data appear overdamped; this may be because of a spin-reorientation transition near 200K, or it may arise from elastic scattering near the laser excitation wavelength. These results agree generally with those in Refs. [14-16]. The $\sigma$ -magnon at 18.2 cm$^{-1}$ varies smoothly from 80K to 280K, but neither its frequency nor its intensity varies monotonically with temperature; instead there is an abrupt increase in intensity at 140 and 200K, and a shift down in frequency by ca. 4 cm$^{-1}$. The variation of frequency shift and intensity of $\gamma$ -magnon at 26.6 cm$^{-1}$ with temperature are presented in Fig 2 respectively. The $\gamma$ -magnon at 26.6 cm$^{-1}$ disappears above 140K and 200K in VH and VV configurations respectively and its frequency is temperature independent from 80K to 200K and shifted only up to 1 cm$^{-1}$. These effects are reproducible on heating and cooling. Their origin is unknown, although other RFeO$_3$ materials are known to have spin reorientation transitions at comparable temperatures(e.g.,84-94K in TmFeO$_3$ [16]).

In a separate paper [10] we will show that zero field cooled (ZFC) and field cooled (FC) magnetization curves in single-crystal bismuth ferrite split below ca. 250K and that the ZFC curves exhibit a sharp cusp at 50K. This suggests a spin-glass behavior at low temperatures with competition between ferromagnetic and antiferromagnetic ordering and may relate to the unexplained anomaly near 200K in the present work.

It is useful to try to relate our data and especially the model used to other descriptions. A rather detailed model of magnon in the spiral structure of BiFeO$_3$ has been made available recently [18] that predicts a series of approximately evenly spaced lines in the Raman effect, resembling the electron Landau levels in a semiconductor under applied magnetic fields. After our work was complete we were alerted to unpublished Raman data [19] which seem in good agreement with that model. Here we would like to relate that model briefly to the one we use, and to the four-dimensional description of Janner and Janssen [20] for incommensurate structures. First, although the experiments and data of [19] agree well for the frequency spacing of the magnon lines, they do not explain why two lines are much more intense than all the others; this is odd because they are not the first lines in either of two sequences. Second, the model does not explain the number of lines - that is, why the sequences stop at six or seven transitions and do not continue. We suggest that there are actually three reasonable models for the data. The model of [18] gives a good description at very low temperatures, where most of the transitions have comparable intensity. However, for an incommensurate spiral structure the four-dimensional representations of Janner and Janssen [20] offer a rigorous alternative. The Janner-Janssen model predicts three modes in [4D] for maximal symmetry. These are labeled R3(00Γ), R3(00Γ)t, and Rc(00Γ). Of these three, the R3(00Γ)t is a transverse mode and the others longitudinal. Thus, using only the incommensurate character of the structure, we anticipate three Raman modes – one transverse and two longitudinal, in accord with the high-intensity modes we see in Fig.2. It would appear that this description is good in the temperature region from ca. 140K to above room temperature.

A further connection emphasizing the fact that only two or three of the many electro-magnon branches have strong Raman intensity is the correspondence with the two sigma and gamma branches in orthoferrites. As emphasized in the introduction, the magnon spectra in orthoferrites consist of two strong lines, and these materials, like $BiFeO_3$ exhibit a spin-reorientation transition. Although $BiFeO_3$ is rhombohedral crystallographically (chemical unit cell) at low T, its magnetic point group is monoclinic, with a very small distortion from orthorhombic. Therefore it resembles the orthorhombic orthoferrites.

Finally, in a separate paper we show that $BiFeO_3$ exhibits spin glass behavior at low temperature with $T_{SG}$ (spin-glass) = 29.4K [10]. In the present context it is important that the spin glass exponent $zv$ = 1.4. This is close to the value 2.0 in the original mean-field Kirkpatrick-Sherrington model [21] and very different from the values ca. 9-10 for Ising models [22]. This seems compatible with the present interpretation of spectra as electro-magnons; these modes are not pure spin waves, and consequently their electric dipole part is Coulombic and long-range, yielding apparently mean-field exponents.

In conclusion, high-resolution Raman data are presented showing two one-magnon branches in bismuth ferrite. The lower branch near 18.4 cm$^{-1}$ is ferromagnetic-like and varies with temperature as an S=5/2 Brillouin function up until 0.44 $T_N$, above which it becomes overdamped or instrumentally unresolved from the elastically scattered laser light. The higher-frequency magnon near 26.6 cm$^{-1}$ is weaker in intensity, rather temperature independent, and resembles the antiferromagnetic mode in orthoferrites. This higher frequency magnons disappears around 140K and 200 in different configuration geometry. Anomalies in magnon frequency and temperature are unexpectedly found at 200K and 140K; these may suggest a spin reorientation, as exists in orthoferrites.

We gratefully acknowledge the financial support from the DOD grant W911NF – 06 – 1 – 0030 and W911NF-06-1-0183.

## *** Figure Captions**

**Fig.1.** Raman spectra from magnon in $BiFeO_3$ single crystal at 80K.

**Fig. 2.**  Temperature dependent electro-magnon spectra of a $BiFeO_3$ single crystal (a) in VV configuration (b) VH configuration.

**Fig. 3.**  Temperature dependent of magnon mode at 18.2 cm[-1] in temperature range between 80 to 280K (a) Integrated Intensity (b) FWHM (c) Experimental correlation between the electromagnon frequency softening of the one-magnon branch at 18.2 cm[-1] and the S=5/2 Brillouin function

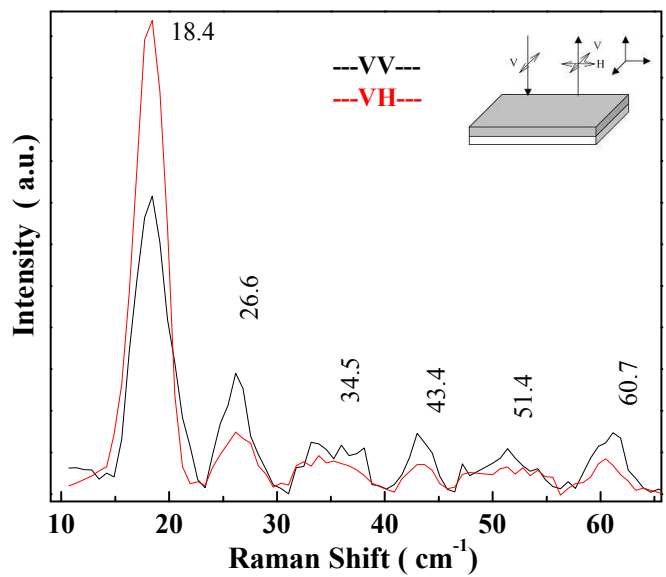

**Figure 1:**

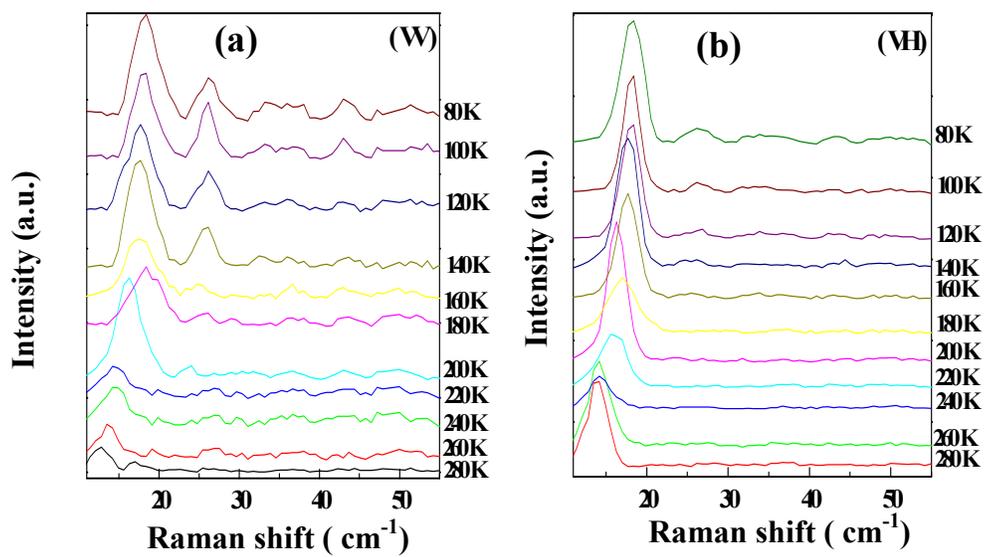

**Figure 2:**

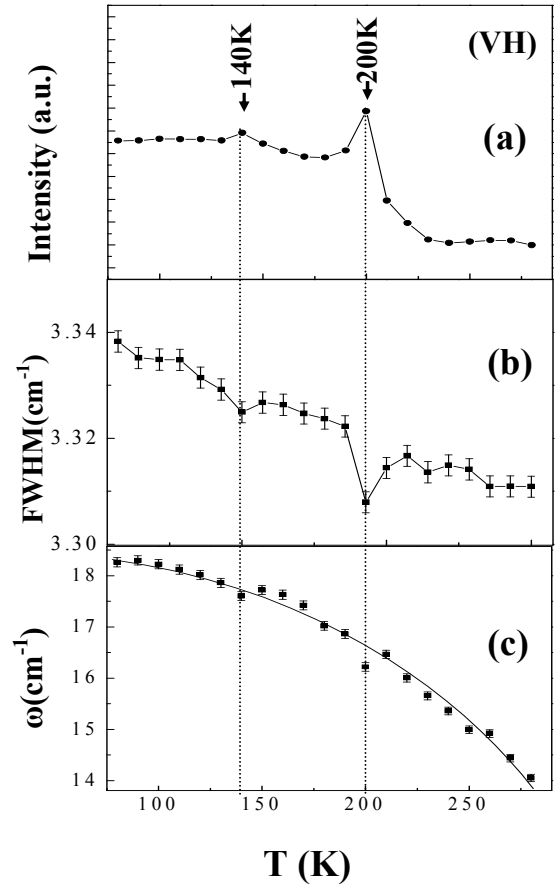

**Figure 3:**